\documentclass[prb,twocolumn,,floatfix]{revtex4}
\usepackage{amsmath}
\usepackage{graphicx}
\usepackage{bm}
\usepackage{amssymb}
\usepackage{color}

\newcommand{\be}{\begin{equation}}
\newcommand{\ee}{\end{equation}}

\def \be{\begin{equation}}
\def \ee{\end{equation}}
\def \ba{\begin{array}}
\def \ea{\end{array}}
\def \bea{\begin{eqnarray}}
\def \eea{\end{eqnarray}}
\def \nn{\nonumber}
\def \half{{1\over 2}}

\def \a{{\alpha}}

\def \b{{\beta}}

\def \nd{{^{\vphantom{\dagger}}}}
\def \yd{^\dagger}

\begin{document}

\title{Rise and fall of hidden string order of lattice bosons}

\author{Erez Berg$^1$, Emanuele G. Dalla Torre$^2$, Thierry Giamarchi$^3$
and Ehud Altman$^2$}
\affiliation{$^1$ Department of Physics, Stanford University, Stanford, CA 94305-4045, USA%
\\
$^2$ Department of Condensed Matter Physics, Weizmann Institute of Science,
Rehovot, 76100, Israel\\
$^3$DPMC-MaNEP, University of Geneva, 24 Quai Ernest-Ansermet, 1211 Geneva,
Switzerland}

\begin{abstract}
We investigate the ground state properties of a newly discovered phase of
one dimensional lattice bosons with extended interactions\cite{DallaTorre}.
The new phase, termed the Haldane Insulator (HI) in analogy with the gapped
phase of spin-1 chains, is characterized by a non local order parameter,
which can only be written as an infinite string in terms of the bosonic
densities. We show that the string order can nevertheless be probed with
physical fields that couple locally, via the effect those fields have on the
quantum phase transitions separating the exotic phase from the conventional
Mott and density wave phases. Using a field theoretical analysis we show
that a perturbation which breaks lattice inversion symmetry gaps the
critical point separating the Mott and Haldane phases and eliminates the
sharp distinction between them. This is remarkable given that neither of
these phases involves broken inversion symmetry. We also investigate the
evolution of the phase diagram with the tunable coupling between parallel
chains in an optical lattice setup. We find that inter-chain tunneling
destroys the direct phase transition between the Mott and Haldane insulators
by establishing an intermediate superfluid phase. On the other hand coupling
the chains only by weak repulsive interactions does not modify the structure
of the phase diagram. The theoretical predictions are confirmed with
numerical calculations using the Density Matrix Renormalization Group (DMRG).

\end{abstract}

\date{\today }
\maketitle

\section{introduction}

Systems of ultracold bosons in optical lattices offer unique opportunities
for studying strongly correlated quantum matter in a highly controllable
environment\cite{BlochReview}. In a sufficiently deep lattice potential the
interactions dominate over the kinetic energy and, at commensurate filling,
can drive a quantum phase transition from the superfluid to the Mott
insulating state. This transition has been observed experimentally\cite%
{Bloch}, and it requires only local (on-site) interactions between
the bosons. Longer range interactions, effective for example in
systems of ultracold polar molecules\cite{molecules} or atoms with
large magnetic dipole moment\cite{dipolar-atoms,dipolar-atoms1} ,
can give rise to even richer behavior. In particular, we have
recently predicted that bosons with sufficiently strong nearest
neighbor or further range repulsion on a one dimensional
lattice 
form a new insulating ground state characterized by hidden topological order
\cite{DallaTorre}.

We termed the new phase a Haldane Insulator (HI) because of the close
analogy with Haldane's gapped phase of integer spin-chains\cite{haldane}.
Both states support a highly non local string order parameter\cite%
{DenNijs,Cirac,Nussinov}. Moreover, the phase transitions from the
Haldane insulating phase to the conventional Mott and Density wave
(DW) insulators, also have
their analogies in anisotropic spin chains \cite%
{Schulz,DenNijs,KennedyTasaki}. This phase can potentially be realized in a
system of cold dipolar atoms or molecules in a one dimensional optical
lattice, where the dipole moment is polarized perpendicular to the chain
direction.

There are, however, essential differences between the spin chains and the
lattice bosons. First, 
the anisotropic spin chains enjoy a global $Z_2$ symmetry associated with
flipping the $S^z$ component of all spins. This would translate to a
particle-hole symmetry about the mean lattice filling, which is clearly
absent in the microscopic bosonic models. Furthermore, the optical lattice
systems are in practice not strictly one dimensional. Some amount of
residual coupling between chains due to tunneling and, in our 
case, also dipolar interactions, is inevitable. More generally, we shall see
that additional external fields, which can be applied to the ultracold Bose
systems, couple in a non trivial way to the string order parameter. The
effects of these perturbations on the phase diagram raise new fundamental
questions on the nature of the non local order.

In this paper we investigate how different perturbations, natural
to bosonic systems, affect the transitions from the Haldane
insulator to the conventional Mott and Density wave phases. We
argue that these directly probe the nature of the non local order
parameter which characterizes the Haldane insulator phase. Using a
bosonization analysis, we demonstrate that the transitions between
the HI and the conventional phases are not sensitive to the broken
particle-hole symmetry in the microscopic Hamiltonian. Such
perturbations also leave the string order parameter intact. On the
other hand we predict that perturbations that break the lattice
inversion symmetry in addition to the particle-hole symmetry,
would eliminate the distinction between the MI and HI phases and
gap out the critical point separating them. Next we investigate
the effect of coupling between two
parallel chains (``ladder" geometry). 
The predictions of the field theoretical analysis are confirmed
with numerical simulations using the Density Matrix
Renormalization Group (DMRG) method.

\begin{figure}[t]
\centering
\includegraphics[width=8cm]{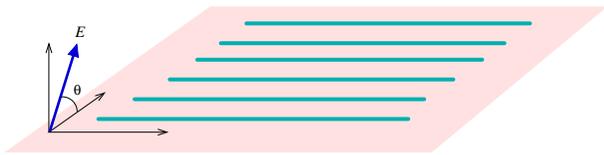}
\caption{(Color online.) Array of one dimensional chains formed by
a two dimensional optical lattice. The polarizing field E is
directed perpendicular to the chains and at an angle
${\protect\theta }$ which can be used to tune the inter-chain
interaction $V_{\perp }$ in (\protect\ref{Hperp}).}
\label{fig:tubes}
\end{figure}

Our starting point for theoretical analysis is the extended Bose Hubbard
model (EBHM) on one or two chains at single site occupation (${\bar{n}}=1$):
\begin{eqnarray}
H_{\alpha } &=&-t\sum_{j}\left( b_{\alpha j}^{\dagger }b_{\alpha j+1}^{%
\vphantom{\dagger}}+h.c.\right)  \notag \\
&&+\frac{U}{2}\sum_{j}n_{\alpha j}\left( n_{\alpha j}-1\right)
+V\sum_{j}n_{\alpha j}n_{\alpha j+1},  \label{EBHM1}
\end{eqnarray}%
where $j$ is the site index and $\alpha=1,2$ is the chain index,
in the case of two chains. The inter-chain coupling in that case
includes a transverse tunneling matrix element and interaction
term:
\begin{equation}
H_{\perp }=-t_{\perp }\sum_{j}(b_{1j}^{\dagger }b{^{\vphantom{\dagger}}}%
_{2j}+H.c.)+V_{\perp }\sum_{j}n_{1j}n_{2j}.  \label{Hperp}
\end{equation}%
In practice the inter-chain tunneling $t_{\perp }$ may be tuned by varying
the depth of the optical lattice potential in the direction perpendicular to
the chains. The inter-chain dipolar interaction can be controlled by
changing the direction of the external polarizing field relative to the
plane of the chains, while keeping it perpendicular to the chain axis. A
schematic setup of this type involving many coupled chains is illustrated in
Fig. \ref{fig:tubes}.
The 
two chain model
\begin{equation}
{H=\sum_{\alpha =1,2}H_{\alpha }+H_{\perp }}  \label{H2}
\end{equation}
serves as a bridge between one and two dimensional geometries, while still
amenable to powerful techniques for treating one dimensional systems, such
as bosonization and the density matrix renormalization group (DMRG). Similar
quasi one-dimensional systems were studied previously in different parameter
regimes: with only on-site interactions\cite{GiamarchiLadder,LuthraLadder}
and with longer-range interactions at half-integer filling\cite{LeeLadder}.
However, the Haldane insulator phase which is the focus of this paper is not
obtained in these regimes.

Recently Anfuso and Rosch noted that the string order characterizing the
Haldane gapped phase of spin-$1$ chains is unstable against inter-chain
antiferromagnetic exchange\cite{Rosch}. The string order in the HI phase is
similarly sensitive to inter-chain tunnel coupling (see appendix \ref%
{app:StringStability}), and the distinction between the HI and MI phases is
lost. What is then the fate of the direct second order transition found
between these two phases when inter-chain tunneling is turned on?
Interestingly we find that direct transition between these two phases is
avoided in the double chain system by the appearance of an intermediate
superfluid phase for arbitrarily small inter-chain tunneling.

Our analysis consists of the following parts. In section \ref{sec:weak1} we
present 
a low energy effective field theory for the single chain, which correctly
captures all three insulating phases MI, HI and DW. This is done with help
of a special bosonization procedure borrowed from work on integer spin chains%
\cite{Schulz}. The bosonized forms for the non local correlations
that characterize the HI and MI phases are given in section
\ref{sec:op}. This framework is used in section \ref{sec:inv} to
study the coupling of the non local order parameters to inversion
symmetry breaking perturbations.
Then in section %
\ref{sec:weak2} we extend the bosonization approach to describe a pair of
weakly coupled chains. We carry out a renormalization group (RG) analysis to
obtain analytic predictions for the phases arising at weak inter-chain
coupling in the vicinity of the quantum critical points. The most
interesting result of the coupling is
the effect of inter-chain tunneling on the
transition between the MI and HI states. This perturbation is highly
relevant at the critical point.
As a result we find that for any finite $t_\perp$ the HI and MI
insulating phases are separated by a superfluid region, whose domain grows
rapidly with increasing $t_\perp$. Finally in section \ref{sec:dmrg} we
confirm the analytic predictions and extend them to stronger inter-chain
coupling using numerical DMRG calculations.

The paper is followed by three appendices. In appendix \ref%
{app:StringStability} we investigate the effect of weak inter-chain coupling
deep in the HI phase within perturbation theory. Although the thermodynamic
HI phase is protected by a gap $\Delta$, we show that the single chain
string order is destroyed even by infinitesimal coupling. In appendix \ref%
{app:bos} we provide an alternative derivation of the effective field theory
using direct bosonization of the particles rather than resorting to an
effective spin-1 model as done in section \ref{sec:weak}. In appendix \ref%
{app:BosString} we construct explicit expressions for the string order
parameter within the effective field theory.

\section{Bosonization}

\label{sec:weak}

\subsection{Continuum limit of a single unperturbed chain}

\label{sec:weak1}

A single chain of interacting bosons (\ref{EBHM1}) at filling ${\bar n}=1$
was studied in Ref. [\onlinecite{DallaTorre}] using DMRG and the existence
of the new Haldane insulator (HI) phase was predicted. The phase diagram in
the two dimensional space of $U/t$ versus $V/t$ is reproduced here in Fig. %
\ref{fig:chain}(a).

The insulating phases and the correct quantum phase transitions separating
them can also be obtained from a field theoretical analysis\cite%
{LeeSchulzBos}, which builds on a direct analogy with the bosonization
procedure developed for integer spin chains\cite%
{TimonenLuther,LutherScalapino,Schulz}. Let us briefly review the derivation
of the effective field theory of a single chain. The first step is an
approximate mapping of the EBHM (\ref{EBHM1}) to an anisotropic spin-1
model. This is done by truncation of the Hilbert space of each site to the
three lowest occupation states $n=0,1,2$\cite{AltmanAuerbach2002,Huber}.
Such a truncation is justified at large $U$ when fluctuations in the site
occupations are strongly suppressed. The effective spin-1 model is
\begin{equation}
H=-t\sum_{j}\left( S_{j}^{+}S_{j+1}^{-}+h.c.\right) +\frac{U}{2}%
\sum_{j}\left( S_{j}^{z}\right) ^{2}+V\sum_{j}S_{j}^{z}S_{j+1}^{z}
\label{H_s}
\end{equation}%
Here $S^{z}=n-\bar{n}$. In general, there are other terms in this
Hamiltonian which break the $S^{z}\rightarrow -S^{z}$, $S^{+}%
\longleftrightarrow S^{-}$ (\textquotedblleft
particle-hole\textquotedblright ) symmetry. However, in the effective
long-wavelength Hamiltonian we will now derive, these terms are irrelevant,
so we will drop them for now. The effect of these terms will be considered
in section \ref{sec:inv}.

One then proceeds by writing each spin 1 variable as a sum of two spin $%
\frac{1}{2}$ variables\cite{TimonenLuther,Schulz}, $%
S_{j}^{z}=s_{1,j}^{z}+s_{2,j}^{z}$. Each spin-$\frac{1}{2}$ chain can now be
mapped to a spinless fermion chain by a Jordan-Wigner transformation. The
fermions are then bosonized in the standard way\cite{GiamarchiBook},
according to $\psi _{\alpha ,R/L}\sim \frac{1}{\sqrt{2\pi a}}e^{i\left(
\theta _{\alpha }\pm \phi _{\alpha }\right) }$, where $a$ is the lattice
spacing, $\alpha =1,2$ corresponds to the two fictitious spin-$1/2$ chains
(not to be confused with the two physical chains that will be considered in
section \ref{sec:weak2}) and $R/L$ corresponds to right and left moving
fermions, respectively. The bosonic fields satisfy the canonical commutation
relations $\left[ \phi _{\alpha }\left( x\right) ,\theta _{\beta }\left(
x^{\prime }\right) \right] =-i\pi \delta _{\alpha ,\beta }\Theta \left(
x^{\prime }-x\right) $, where $\Theta \left( x\right) $ is a Heaviside step
function. With these conventions\cite{ShankarBosonization}, the spin-$1/2$
operators have the following bosonized form:
\begin{equation}
s_{\alpha }^{z}=\frac{a}{\pi }\partial _{x}\phi _{\alpha
}+\frac{\left( -1\right) ^{\frac{x}{a}}}{\pi }\sin \left( 2\phi
_{\alpha }\right) \text{.}\label{sz_bos}
\end{equation}

The bosonized Hamiltonian is conveniently written in terms of symmetric ($+$%
) and antisymmetric $(-)$ combinations of the fields $\phi _{\alpha }$ and $%
\theta _{\alpha }$:
\begin{equation}
H=H_{+}+H_{-}+H_{+-}.  \label{bosH1}
\end{equation}%
The three terms are given explicitly by:
\begin{eqnarray}
H_{+} &=&\frac{u_{+}}{2\pi }\int dx\left[ K_{+}\left( \partial _{x}\theta
_{+}\right) ^{2}+\frac{1}{K_{+}}\left( \partial _{x}\phi _{+}\right) ^{2}%
\right]  \notag \\
&&+\int dx\frac{g_{1}}{\left( \pi a\right) ^{2}}\cos \left( 2\phi _{+}\right)
\label{Hp} \\
H_{-} &=&\frac{u_{-}}{2\pi }\int dx\left[ K_{-}\left( \partial _{x}\theta
_{-}\right) ^{2}+\frac{1}{K_{-}}\left( \partial _{x}\phi _{-}\right) ^{2}%
\right]  \notag \\
&&+\int dx\left[ \frac{g_{2}}{\left( \pi a\right) ^{2}}\cos \left( 2\phi
_{-}\right) +\frac{g_{3}}{(\pi a)^{2}}\cos \left( 2\theta _{-}\right) \right]
\label{Hm}
\end{eqnarray}%
and
\begin{equation}
H_{+-}=\int dx\frac{g_{4}}{\left( \pi a\right) ^{2}}\cos \left( 2\phi
_{+}\right) \cos \left( 2\phi _{-}\right) ,  \label{singlechain2}
\end{equation}%
where $\phi _{\pm }=\phi _{1}\pm \phi _{2}$ 
and $\theta _{\pm }=\left( \theta _{1}\pm \theta _{2}\right) /2$.
(Note that the fields $\phi _{+}$ and $\phi _{-}$ correspond to $\sqrt{2}%
\psi _{1}$ and $\sqrt{2}\psi _{2}$ in Ref. [\onlinecite{Schulz}],
respectively.) The naive continuum limit gives the following estimates for
bare values of the coupling constants:
\begin{eqnarray}
u_{+} &=&ta\sqrt{1+\frac{U+6V}{\pi t}},~~K_{+}=\frac{2}{\sqrt{1+\frac{U+6V}{%
\pi t}}}  \notag \\
u_{-} &=&ta\sqrt{1-\frac{U-2V}{\pi t}},~~K_{-}=\frac{2}{\sqrt{1-\frac{U-2V}{%
\pi t}}}  \notag \\
g_{1} &=&-g_{2}=\frac{\left( 2V-U\right) a}{2},~g_{3}=-t\pi a,~g_{4}=Va
\label{weak_params}
\end{eqnarray}

To study the physical correlation functions we need to express the second
quantized Bose operators on the lattice and the local site occupation in
terms of the continuum fields. These relations are of the form
\begin{eqnarray}
\frac{{b}\left( x\right) }{\sqrt{a}} &=&\frac{A}{\sqrt{2\pi a}}e^{i\theta
_{+}}\left[ \cos \left( \theta _{-}\right) +...\right]  \label{b_bos} \\
\frac{n\left( x\right) }{a} &=&\frac{1}{\pi }\partial _{x}\phi _{+}+B\frac{%
\left( -1\right) ^{\frac{x}{a}}}{\pi a}\sin \left( \phi _{+}\right) \cos
\left( \phi _{-}\right) +...,  \notag
\end{eqnarray}%
where $A$ and $B$ are non-universal constants and only the most relevant
terms are shown. In general, the expansion of these operators also contains
less relevant (sub-leading) terms.

The Hamiltonian $H_{+}$ of the symmetric degrees of freedom is
similar to the usual low energy description of lattice bosons. The
Umklapp term parameterized by $g_{1}$ is relevant for $K_{+}<2$.
Here we consider the insulating phases, for which $K_{+}$ is
indeed below this critical value\footnote{The superfluid phase,
which is realized for small $U/t, V/t$, is not captured by this
approach, due to the truncation to $n=0,1,2$ on each site. See
Appendix \ref{app:bos} for a bosonization scheme that captures the
transition to the superfluid phase correctly.}. The main
difference from the standard bosonization of lattice bosons is
that the parameter $g_{1}$ changes sign at $2V=U$. This change of
sign marks a quantum phase transition from the MI, for which $\phi
_{+}$ is localized around zero, to another gapped phase in which
it is localized near $\pm \frac{\pi }{2}$. As we shall see below,
the latter turns out to be the HI phase and is characterized by
string order. On the critical line, the system is described by a
Luttinger liquid with power law correlations that depend on the
non-universal parameter $\frac{1}{2}<K_{+}<2$. If $K_{+}$ is
driven below $\frac{1}{2}$, the critical line becomes unstable,
since the sub-leading term $\cos(4\phi_+)$ (which was omitted in
Eq. (\ref{Hp})) becomes relevant. This possibility will be
discussed further in Sec. \ref{sec:inv}. Note that the odd sector
described by $H_{-}$ is not critical. The term $g_{3}$ in
(\ref{Hm}) is relevant on both sides of the transition, so that
the field ${\theta }_{-}$ is localized and remains massive at the
critical point.

On the other hand, the transition from the HI\ to the DW\ phase is
controlled by the odd sector. It occurs when the term $\cos \left( 2\phi
_{-}\right) $ becomes more relevant than the dual term $\cos \left( 2\theta
_{-}\right) $, that is when $K_{-}<1$. According to the naive continuum
limit the critical line is given by $V-2U=3\pi t$. This transition is of the
Ising universality class\cite{Schulz,numerics_xxz}. This can also bee seen
by refermionization of the Hamiltonian $H_{-}$ at the critical point $%
K_{-}=1 $ and $g_{2}=g_{3}$, which yields precisely the massless Majorana
fermions of the $1+1$ dimensional Ising model at criticality.

We remark that the precise values of the coupling constants as a function of
the microscopic parameters are not simple to derive. The field theory (\ref%
{bosH1}) should be used to obtain the universal behavior of the system near
the critical points which describe the quantum phase transitions, not to
find the precise location of phase boundaries.

\subsection{The string and parity order parameters}

\label{sec:op}

\begin{figure}[b]
\centering
\includegraphics[width=9cm]{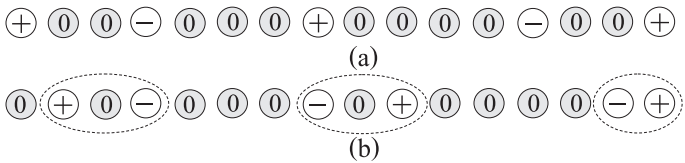}
\caption{Typical configurations in the HI (a) and MI (b) ground states. The
numbers represent $\protect\delta n$ (the deviation of the local occupation
from the average density). The HI can be described as a charge ordered $%
+,-,+,...$ state with an undetermined number of 0 sites between each $+$ and
$-$. The MI is a dilute gas of particle-hole pairs (indicated by the dotted
line).}
\label{fig:MI_HI_pic}
\end{figure}

What is the physical distinction between the different insulating phases? It
was shown in Ref. [\onlinecite{DallaTorre}] that the HI phase of bosons is
characterized by a non decaying string correlation function
\begin{equation}
\mathcal{O}_{S}^{2}\equiv \underset{\left\vert i-j\right\vert \rightarrow
\infty }{\lim }\left\langle \delta n_{i}\exp \left( i\pi \sum_{i\leq
k<j}\delta n_{k}\right) \delta n_{j}\right\rangle .  \label{Cs}
\end{equation}%
Here ${\delta }n_{i}\equiv n_{i}-\bar{n}$ is the deviation from the average
(integer) filling $\bar{n}$. In section \ref{sec:dmrg} we show that the MI
phase is characterized by a different non local correlation function, which
will be termed the \textquotedblleft parity" correlation function
\begin{equation}
\mathcal{O}_{P}^{2}\equiv \underset{\left\vert i-j\right\vert \rightarrow
\infty }{\lim }\left\langle \exp \left( i\pi \sum_{i\leq k<j}\delta
n_{k}\right) \right\rangle .  \label{Cp}
\end{equation}%
For the sake of convenience, we define the parity and string operators
\begin{equation}
\hat{\mathcal{O}}_{P}\left( j\right) =\exp \left( i\pi \sum_{k<j}\delta
n_{k}\right)   \label{Op}
\end{equation}%
\begin{equation}
\hat{\mathcal{O}}_{S}\left( j\right) =\hat{\mathcal{O}}_{P}\left( j\right)
\delta n_{j},  \label{Os}
\end{equation}%
such that $\mathcal{O}_{P}^{2}=\lim_{\left\vert i-j\right\vert \rightarrow
\infty }\left\langle \hat{\mathcal{O}}_{P}\left( i\right) \hat{\mathcal{O}}%
_{P}\left( j\right) \right\rangle $ and $\mathcal{O}_{S}^{2}=\lim_{\left%
\vert i-j\right\vert \rightarrow \infty }\left\langle \hat{\mathcal{O}}%
_{S}\left( i\right) \hat{\mathcal{O}}_{S}\left( j\right) \right\rangle $.

A simple interpretation of these correlations is illustrated in Fig. \ref%
{fig:MI_HI_pic}. In both phases the ground state consists of
configurations with most sites having precisely the average
occupation, but also some particle and hole fluctuations. The
parity $\mathcal{O}_{P}$ order in the MI phase implies that the
fluctuations in this ground state are bound particle-hole pairs.
On the other hand, the string order in the HI phase implies that a
\textquotedblleft renormalized" chain with all non fluctuating
${\delta }n_{i}=0$ sites taken away would have density wave order
(particle and hole fluctuations alternating along the chain).
Physically, this can be understood as a compromises between the
kinetic energy term in Eq. (\ref{EBHM1}), which prefers maximum
delocalization of particles and holes, and the nearest neighbor
interaction term, which is minimized when particles and holes are
neighbors. Indeed, the HI phase is realized at intermediate values
of $V/t$.

Within the effective field theory the HI and MI seem to differ only in the
expectation value of the field $\phi _{+}$ which is pinned in each of these
phases. It is interesting to relate this distinction to the string and
parity correlations that characterize the two phases. Since both order
parameters contain the factor $\exp \left( i\pi \sum_{i<k<j}\delta
n_{k}\right) $, we may naively expect their bosonized forms to contain $\exp %
\left[ i\phi _{+}\left( x\right) \right] $, since $\sum_{i<k<j}\delta
n_{k}\rightarrow \frac{1}{\pi }\int_{x_{i}}^{x_{j}}dx\partial _{x}\phi _{+}=%
\frac{1}{\pi }\left[ \phi _{+}\left( x_{j}\right) -\phi _{+}\left(
x_{i}\right) \right] $. However, the exponential should be symmetrized
carefully to obtain a hermitian operator. In Appendix \ref{app:BosString},
we argue for the following forms of $\hat{\mathcal{O}}_{S}$ and $\hat{%
\mathcal{O}}_{P}$ in the bosonized theory:
\begin{equation}
\hat{\mathcal{O}}_{S}(x)\sim \sin \left( \phi _{+}\left( x\right) \right)
\label{Os_bos}
\end{equation}%
\begin{equation}
\hat{\mathcal{O}}_{P}(x)\sim \cos \left( \phi _{+}\left( x\right) \right)
\label{Op_bos}
\end{equation}%
Here the form of (\ref{Os_bos}) was postulated on the basis of symmetry. For
a more microscopic derivation see Refs. [\onlinecite{GiamarchiString},%
\onlinecite{Nakamura}].

The above expressions for the string and parity correlations are consistent
with the respective phases derived from Eq. (\ref{Hp},\ref{Hm}). The MI
phase corresponds to $g_{1}<0$, which implies a non vanishing expectation
value of the parity operator $\left\langle \cos \left( \phi _{+}\left(
x\right) \right) \right\rangle \neq 0$. The fact that $\phi _{+}$ is locked
to $0$ or $\pi $ in this phase is consistent with the particle density being
concentrated on the lattice sites, as in the cartoon product state ${%
\,|\,\Psi _{MI}\,\rangle \,}\sim \prod_{j}b_{j}^{\dagger }{\,|\,0\,\rangle \,%
}$. In the HI\ phase on the other hand $g_{1}>0$, so that$\left\langle \sin
\left( \phi _{+}\left( x\right) \right) \right\rangle \neq 0$ and therefore
non vanishing string order. Here $\phi _{+}$ is locked to $\pi /2$ or $3\pi
/2$, which implies a shift of the particles by half a lattice constant
compared with the MI, that is, the density is centered on the links rather
than the lattice sites. This is captured by the cartoon wave function ${%
\,|\,\Psi _{HI}\,\rangle \,}\sim \prod_{j}(b_{j}^{\dagger }+b_{j+1}^{\dagger
}){\,|\,0\,\rangle \,}$, which is closely analogous to the AKLT state that
describes a valence bond solid in spin-1 chains.


\subsection{Coupling to symmetry breaking perturbations}

\label{sec:inv}

A realization of the phases described above with ultracold atoms would open
up new ways to probe the nature of string and parity orders by how they
react to different perturbations. Interestingly, we shall see that in spite
of being highly non local, the string and parity operators nevertheless
couple in interesting ways to local symmetry breaking fields.

As compared to the spin chain model [Eq. (\ref{H_s})],
an inherent broken symmetry in the microscopic Hamiltonian (\ref{EBHM1}) is
the absence of particle hole symmetry ${\delta }n_{i}\rightarrow -{\delta }%
n_{i}$. The spin model does enjoy the analogous symmetry associated with
rotation by $\pi $ around the $y$ axis, which takes $S^{z}\rightarrow -S^{z}$
and $S^{x}\rightarrow -S^{x}$. Nevertheless both systems are described by
the same low energy field theory (\ref{bosH1}). Terms that break just the
particle hole symmetry, such as $\partial _{x}\phi (\partial _{x}{\theta }%
)^{2}$, are irrelevant at the $HI$-$MI$ critical point, and do not change
either the parity ${\langle \cos }\left( \phi _{+}\right) {\rangle }$ or
string ${\langle \sin \left( \phi _{+}\right) \rangle }$ order parameters in
the insulating phases. What are the minimal perturbations that eliminate the
sharp distinction between these two phases and gap out the critical point
separating them?

Consider the following perturbation of the Hamiltonian (\ref{Hp}):
\begin{eqnarray}
H_{+}+\lambda {\hat{P}} &=&H_{+}+\lambda \int {\frac{dx}{(\pi a)^{2}}}\sin
(2\phi _{+})  \notag \\
&=&H_{LL}+{\tilde{g}}\int {\frac{dx}{(\pi a)^{2}}}\cos (2\phi _{+}+\chi ).
\label{Hdipole}
\end{eqnarray}%
Here $H_{LL}$ is the critical Luttinger liquid theory at the transition and $%
\chi =\arctan (\lambda /g_{1})$ and ${\tilde{g}}=\sqrt{g_{1}^{2}+\lambda ^{2}%
}$. Clearly $\hat{P}$ is a relevant perturbation at the critical point,
leading to a gapped state even when the parameter $g_{1}$ crosses zero,
where the critical point would have been.

Note that the operator $\hat{P}$ breaks both the particle hole and lattice
inversion symmetries. This is seen for example in the representation of the
single chain as two coupled fermionic chains (as in Ref. [\onlinecite{Schulz}%
]). Written in terms of the fermions the perturbation is
\begin{equation}
\sin \left( 2\phi _{+}\right) \sim i\psi _{R,1}^{\dagger }\psi
_{R,2}^{\dagger }\psi _{L,1}^{\vphantom{\dagger}}\psi _{L,2}^{%
\vphantom{\dagger}}+H.c.,
\end{equation}%
where $\psi _{R,\alpha }^{\vphantom{\dagger}}$ and $\psi _{L,\alpha }^{%
\vphantom{\dagger}}$ for $\alpha =1,2$ are the right and left
moving fermions in the two fermionic chains, respectively. (Note
that the two chains labelled by $\alpha$ are the two fictitious
spin-1/2 chains discussed in the paragraph preceding Eq.
(\ref{sz_bos}).) This term is clearly odd
under inversion (which changes $\psi _{R,\alpha }^{\vphantom{\dagger}%
}\longleftrightarrow \psi _{L,\alpha }^{\vphantom{\dagger}}$) and
particle-hole ($\psi _{R/L,\alpha }^{\dagger }\longleftrightarrow
\psi _{R/L,\alpha }^{\vphantom{\dagger}}$) transformations, and
even under time-reversal.

The perturbation $\hat{P}$ can be viewed as a local polarizing
field that induces a small ``dipole" moment on every lattice site.
(By ``dipole'' we mean $\int{xn(x)dx}$, i.e. the electric dipole
moment that we would get if the particles where charged. It is not
related to the real moment of the dipolar atoms.) It
is clear from the second line of (\ref{Hdipole}) that the term $-\tilde{g}%
\cos (2\phi _{+}+\chi )$ acts to lock the field $\phi _{+}$ to the values $%
\phi _{+}\approx -\chi /2$ or $\phi _{+}\approx \pi -\chi /2$. The density
concentration is shifted accordingly by $\sim (\chi /2\pi )a$ away from the
lattice points, resulting in a local dipole moment at these points. The
shift of $-\chi /2$ also implies that both parity ${\langle \hat{\mathcal{O}}%
_{P}\rangle }$ and string ${\langle \hat{\mathcal{O}}_{S}\rangle }$ order
parameters gain a finite expectation value in the presence of this
perturbation. Hence the sharp distinction between the Mott and Haldane
insulating phases is lost. It is interesting to note in this regard the
similar effect of a symmetry breaking field on a conventional quantum phase
transitions involving spontaneous breaking of that symmetry. The field gaps
out the critical point while inducing a non vanishing order parameter in the
disordered phase. It is remarkable that breaking inversion symmetry has this
effect in the HI-MI transition despite the fact that neither phase has
broken lattice inversion symmetry (at least in an infinite system or in a
finite system with periodic boundary conditions). \cite{FootnoteFinite}

We note that the converse is also true: a finite expectation value of both
parity and string orders immediately entails broken inversion symmetry in
the ground state. This is seen by writing the local symmetry breaking field
as a product of the two non local order parameters $\sin (2\phi _{+})=2\sin
(\phi _{+})\cos (\phi _{+})$. 

The connection between the string and parity correlations and the breaking
of inversion symmetry can also be understood from the microscopic viewpoint
sketched in Fig. \ref{fig:MI_HI_pic}. A non-zero expectation value of the
parity operator ${\langle \hat{\mathcal{O}}_P\rangle}$ is associated with
pairing of particle and hole fluctuations in the ground state, while ${%
\langle \hat{\mathcal{O}}_S\rangle}\ne 0$ corresponds to alternate ordering
of the particle and hole fluctuations. Having both implies organization of
the particle hole pairs in the form of ordered dipoles. A cartoon
wave-function capturing this mixed phase is given by:
\begin{equation}
{\,|\,\Psi_d\,\rangle\,}=\prod_i \left[(1+d)b^\dagger_i+(1-d)b^\dagger_{i+1}%
\right]{\,|\,0\,\rangle\,} \label{psi_d}
\end{equation}
Changing $d$ between $0$ and $1$ facilitates a continuous connection between
the MI and HI phases.

In an optical lattice system it is relatively easy to apply a perturbation
that breaks lattice inversion symmetry. A second laser with double the
wavelength can be used to produce a lattice of asymmetric double wells\cite%
{DoubleWells1,DoubleWells2}. Terms such as $\sin(2\phi_+)$ are then allowed
by symmetry and will therefore be imminently generated in the effective
field theory.

An intriguing question that naturally arises is whether a phase
described by a wave-function such as Eq. (\ref{psi_d}) can occur
by spontaneous breaking of inversion symmetry. In fact, there is a
natural route by which such symmetry breaking can take place. As
discussed in Sec. \ref{sec:weak1}, the critical line between the
MI and HI phases is stable when $K_+>\frac{1}{2}$. For $K_+<\half$
sub-leading terms, such as $\cos(4\phi_+)$, become relevant and
open a gap that would stabilize a new phase. In particular, if the
coefficient of the $\cos(4\phi_+)$ term is positive, the $\phi_+$
field is locked to $\pm\pi/4$, which implies spontaneously broken
inversion symmetry. In this case, the particles (or density
maxima) are effectively pinned to a point which is either $1/4$
lattice spacing to the left or to the right of lattice sites. What
microscopic interactions are required to stabilize such a phase is
an open question. It can only occur if the value of the Luttinger
parameter $K_+$ goes below $\half$ before the transition to the DW
terminates the critical line separating the MI and HI phases. This
point will be investigated in a later study.

A completely different type of perturbation that 
is difficult to realize in cold atom systems, but is nevertheless worth
mentioning, 
is one that breaks the $U(1)$ symmetry associated with number conservation.
Such a perturbation can be added to the continuum theory (\ref{Hp}) as a
term of the form $\lambda \int dx\cos (\theta _{+})$. In the spin chain case
[Eq. (\ref{H_s})], this term corresponds to a transverse magnetic field.
Clearly, this perturbation is relevant and would gap out the critical point.
Inside the HI and MI this term produces a finite density of kinks in the
field $\phi _{+}$ and therefore leads to vanishing of both expectation
values ${\langle \cos (\phi _{+})\rangle }$ and ${\langle \sin (\phi
_{+})\rangle }$. Again the sharp distinction between the HI and MI phases is
lost.

\subsection{Continuum limit of weakly coupled chains}

\label{sec:weak2} We move on to discuss a system of two coupled EBHM chains
described by the microscopic Hamiltonian (\ref{H2}). The starting point for
a low energy description at weak inter-chain coupling are the decoupled
single chain field theories (\ref{bosH1}) which involve four bosonic fields,
$\phi _{\alpha ,\pm }$ (with $\alpha =1,2$ a chain index) and their
canonical conjugate fields ${\theta}_{{\alpha},\pm}$.

The continuum limit of the inter-chain Hamiltonian (\ref{Hperp}) is now
conveniently written in terms of symmetric and anti-symmetric combinations
of the \emph{even }fields of each chain: $\Phi _{\pm }=\left( \phi _{1,+}\pm
\phi _{2,+}\right) /\sqrt{2}$, and $\Theta _{\pm }=\left( \theta _{1,+}\pm
\theta _{2,+}\right) /\sqrt{2}$:
\begin{eqnarray}
H_{V_{\perp }} &=&\int dx{\LARGE \{}\frac{g_{5}}{2\pi ^{2}}\left[ \left(
\partial _{x}\Phi _{+}\right) ^{2}-\left( \partial _{x}\Phi _{-}\right) ^{2}%
\right]  \notag \\
&&+\frac{g_{6}}{\left( \pi a\right) ^{2}}\left( \cos \left( \sqrt{2}\Phi
_{-}\right) -\cos \left( \sqrt{2}\Phi _{+}\right) \right)  \notag \\
&&\times \cos \left( \phi _{1,-}\right) \cos \left( \phi _{2,-}\right) +...%
{\LARGE \}}  \label{Hvperp}
\end{eqnarray}%
\begin{eqnarray}
H_{t_{\perp }} &=&\int dx\frac{g_{7}}{(\pi a)^{2}}\cos \left( \sqrt{2}\Theta
_{-}\right)  \notag \\
&&\times \left[ \cos \left( \theta _{1,-}\right) \cos \left( \theta
_{2,-}\right) +...\right]  \label{Htperp}
\end{eqnarray}%
\begin{figure}[t]
\centering
\includegraphics[width=8cm]{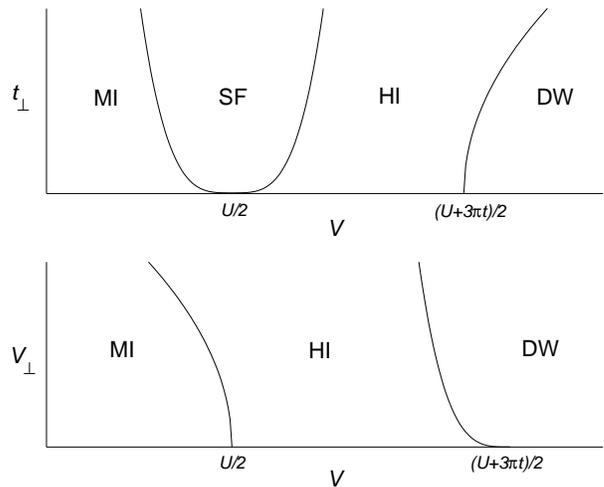}
\caption{Evolution of the phase diagram with inter-chain coupling
predicted
from bosonization. (a) Plane of the nearest neighbor interaction $V$ versus $%
t_{\perp }$ (inter-chain tunneling) at fixed on-site interaction $U$ and no
inter-chain repulsion. (b) In the plane of $V$ versus $V_{\perp }$
(inter-chain repulsion) at $t_\perp=0$ and fixed $U$.}
\label{fig:bos_phase}
\end{figure}

where the bare values of the parameters are $g_{5}=V_{\perp }a$, $%
g_{6}=V_{\perp }aB^{2}$ and $g_{7}=-t_{\perp }\pi a A^{2}$. (Here $A$, $B$
are the non-universal constants that appear in Eq. (\ref{b_bos}).) We shall
analyze how these terms affect the low energy physics of the decoupled
chains (\ref{bosH1}) in the limit of weak $t_{\perp }$ and $V_{\perp }$.

\emph{The HI phase --} The simplest case to consider is the case where the
decoupled chains are in the HI phase. Because of the finite gap in each
chain's spectrum, any inter-chain coupling term is irrelevant in this case,
which implies that the chains remain essentially decoupled at weak coupling.
Nevertheless, since the string order parameter is a non local object in
terms of either the density or the phase field it is destroyed by
infinitesimal inter-chain tunneling. In appendix \ref{app:StringStability}
we show that the string correlations decay exponentially with a correlation
length that scales as $\left( \Delta /t_{\perp }\right) ^{2}$, where $\Delta
$ is the gap (see also Ref. [\onlinecite{Rosch}]). It is easy to understand
this effect by considering the low energy theory (\ref{bosH1}) with the
perturbation (\ref{Htperp}). In the HI phase of a single chain the the field
$\phi _{+}$ is localized near $\pm \pi /2$, so that the string order
parameter ${\langle \sin (\phi _{+})\rangle }$ is fixed around $\pm 1$.
However, the perturbation $t_{\perp }\cos \left( \sqrt{2}\Theta _{-}\right) $%
, which hops a particle from one chain to the other, induces a kink in the
string order parameter of each chain. Therefore at any finite coupling
strength this perturbation destroys the string order by creating a finite
density $\propto \left( t_{\perp }/\Delta \right) ^{2}$ of kinks in the $%
Z_{2}$ order parameter.

We note that the product of the string operators on the two chains remains
invariant to the inter-chain tunneling because this perturbation always
creates a kink on one chain and an anti-kink on the other. It is therefore
tempting to define the product $O_{SL}=\sin (\phi _{+,1})\sin (\phi _{+,2})$
as a generalized string order parameter for the ladder. Such an operator
still has a non-zero expectation value in the HI\ phase when inter-chain
hopping is introduced. However, it gains a finite expectation value also in
the MI phase in the presence of a finite $t_{\perp }$ and therefore cannot
serve to distinguish these two phases. To see this we consider the decoupled
chains in the MI phase $\phi _{+,{\alpha }}\approx 0$. The action of the
inter-chain tunneling operator at point $x$ produces a kink-anti kink pair
corresponding to a change of $\phi _{+,1}$ from $0$ to $\pi $ at $x$ and of $%
\phi _{+,2}$ from $0$ to $-\pi $. At the point $x$ the values of the fields
are $\phi _{+,1}\approx \pi /2$ and $\phi _{+,1}\approx -\pi /2$, therefore
such a point contributes $-1$ to the product string operator. We therefore
expect that in the presence of finite coupling $t_{\perp }$ the product
string operator would gain an expectation value even in the MI phase, which
is of the order of the kink density $\left( t_{\perp }/\Delta \right) ^{2}$.

It is natural to ask what is the fate of the quantum phase transition from
the MI to HI phase in the two chain system where we cannot define a clear
cut distinction between the two phase.

\emph{HI to MI transition --} The critical theory for this transition is a
Luttinger liquid described by $H_{+}$, Eq. (\ref{Hp}), with $g_{1}=0$. In
this region, $H_{-}$ is gapped, with the $\cos \left( 2\theta _{\alpha
,-}\right) $ term more relevant than the $\cos \left( 2\phi _{\alpha
,-}\right) $ term. We may therefore assume that the $\theta _{\alpha ,-}$
fields are pinned at the minimum of the cosine potential. The effective low
energy Hamiltonian in the $\Phi _{\pm }$ sector is written as follows:

\begin{eqnarray}
H_{MI\rightarrow HI} &=&\sum_{\pm }\int dx\frac{\tilde{u}_{\pm }}{2}\left[
\tilde{K}_{\pm }\left( \partial _{x}\Theta _{\pm }\right) ^{2}+\frac{1}{%
\tilde{K}_{\pm }}\left( \partial _{x}\Phi _{\pm }\right) ^{2}\right]  \notag
\\
&&+\int dx\frac{g_{1}}{\left( \pi a\right) ^{2}}\cos \left( \sqrt{2}\Phi
_{+}\right) \cos \left( \sqrt{2}\Phi _{-}\right)  \notag \\
&&+g_{7}\int dx\frac{1}{(\pi a)^{2}}\cos \left( \sqrt{2}\Theta
_{-}\right) , \label{H_MI_HI}
\end{eqnarray}%
where the bare values are $\tilde{u}_{\pm }=u_{\pm }\sqrt{1\pm \frac{%
2g_{5}K_{+}}{\pi u_{+}}}$, $\tilde{K}_{\pm }=K_{+}/\sqrt{1\pm \frac{%
2g_{5}K_{+}}{\pi u_{+}}}$, and $g_{7}\propto t_{\perp }$. Here we have
replaced $\cos \left( \theta _{\alpha ,-}\right) $ by its non-zero
expectation value. Note also that since the fields $\theta _{\alpha ,-}$ are
pinned, the $g_{v_{\perp }}^{\left( 2\right) }$ term in Eq. (\ref{Hvperp})
(which contains $\cos \left( \phi _{\alpha ,-}\right) $) is strongly
irrelevant at this critical point, and therefore it was omitted in Eq. (\ref%
{H_MI_HI}). We will, however, consider its effect on the phase diagram in
what follows.

The Hamiltonian (\ref{H_MI_HI}) is identical to the effective Hamiltonian
derived in Ref. [\onlinecite{GiamarchiLadder}] for a system of two
Bose-Hubbard chains coupled by an inter-chain hopping term. The only
difference is that here, due to the extended interaction terms, a wider
regime of parameters is accessible in Eq. (\ref{H_MI_HI}). Of particular
interest is the transition point from the HI to the MI phase, where $g_1=0$
(which was only possible in the non-interacting limit in Ref. [%
\onlinecite{GiamarchiLadder}]).

For completeness, we will now review briefly the RG analysis of $%
H_{MI\rightarrow HI}$ in Eq. (\ref{H_MI_HI}), following Ref [%
\onlinecite{GiamarchiLadder}]. The leading order RG flow equations are:
\begin{eqnarray}
\frac{dg_{1}}{d\ell } &=&\left( 2-\frac{\tilde{K}_{+}}{2}-\frac{\tilde{K}_{-}%
}{2}\right) g_{1}  \notag \\
\frac{dg_{7}}{d\ell } &=&\left( 2-\frac{1}{2\tilde{K}_{-}}\right) g_{7}
\notag \\
\frac{d\tilde{K}_{+}}{d\ell } &=&-\frac{K_{+}^{2}g_{1}^{2}}{16\pi ^{2}}
\notag \\
\frac{d\tilde{K}_{-}}{d\ell } &=&-\frac{K_{+}^{2}g_{1}^{2}}{16\pi ^{2}}+%
\frac{g_{7}^{2}}{8\pi ^{2}}  \label{RG1}
\end{eqnarray}%
The fate of the system is determined by the competition between the $g_{7}$
and $g_{1}$ terms, which contain the dual cosine terms $\cos \left( \sqrt{2}%
\Theta _{-}\right) $ and $\cos \left( \sqrt{2}\Phi _{-}\right) $,
respectively. If $g_{1}$ dominates, then the system will be in a HI-like
phase (for $g_{1}>0$) or an MI-like phase (for $g_{1}<0$). If $g_{7}$
dominates, then a new phase is stabilized. Note that at the $g_{1}=0$
critical point, the $g_{7}$ term is relevant for $\tilde{K}_{-}>\frac{1}{8}$.

We can use the flow equations (\ref{RG1}) to determine the form of the phase
boundaries that separate the new $\tilde{g}_{t_{\perp }}$ dominated phase
from the HI and MI phases. Imagine that we start from a point on the
critical line between the MI or HI and the large $\tilde{g}_{t_{\perp }}$
phases, $(\tilde{g}_{t_{\perp },c},g_{1,c})$, where both $\tilde{g}%
_{t_{\perp },c}$ and $g_{1,c}$ are small. Integrating the leading order flow
equations, we find that this point is mapped to the point $\left( t_{\perp
,c}(\Lambda _{0}/\Lambda )^{2-1/\left( 2\tilde{K}_{-}\right)
},g_{1,c}(\Lambda _{0}/\Lambda )^{2-\tilde{K}_{+}/2-\tilde{K}_{-}/2}\right) $%
, where $\Lambda _{0}/\Lambda >1$ is the RG scaling factor. Clearly, this
point must also be on the critical line. Therefore the critical line must be
of the form\cite{GiamarchiLadder} $t_{\perp ,c}\propto \left\vert
g_{1,c}\right\vert ^{\alpha }$ where $\alpha =\frac{2-1/\left( 2\tilde{K}%
_{+}\right) }{2-\tilde{K}_{+}/2-\tilde{K}_{-}/2}$. For weak to intermediate
interactions, this gives $\alpha >1$.

To elucidate the nature of the new phase that forms between the Mott and
Haldane insulators for non-vanishing $t_{\perp }$, we consider the possible
sub-leading interactions that may become relevant in this phase. The single
chain even sector Hamiltonian $H_{+}$ (Eq. (\ref{Hp})) can contain higher
harmonics of the form $\cos \left( 2n\phi _{\alpha ,+}\right) $ with $n>1$.
However, when expanded in terms of $\Phi _{\pm }=\left( \phi _{1,+}\pm \phi
_{2,+}\right) /\sqrt{2}$, these terms are seen to be irrelevant since they
always contain the odd field $\Phi _{-}$, whose conjugate $\Theta _{-}$ is
pinned. The term $\cos \left( \sqrt{8}\Phi _{+}\right) $ can also be
generated to higher orders in the bare couplings. (For example, such a term
is generated during the RG flow to second order in $g_{1}$.) This term has a
scaling dimension of $2\tilde{K}_{+}$, therefore it is relevant for $\tilde{K%
}_{+}<1$. This represents a strong interaction, therefore one may still
expect that for small to moderate $U$, $V$ and inter-chain coupling, this
term would be irrelevant. In that case, the intermediate finite $t_{\perp }$
phase is gapless, and characterized by power law correlations. Using the
bosonized expression for the boson creation operator [Eq. (\ref{b_bos})],
the off-diagonal correlation function in this phase is
\begin{equation}
\langle b_{\alpha }(x)^{\dagger }b_{\beta
}^{\vphantom{\dagger}}(0)\rangle \sim \langle e^{i\Theta
_{+}\left( x\right) /\sqrt{2}}e^{-i\Theta _{+}( 0)
/\sqrt{2}}\rangle \sim \frac{1}{\vert x\vert
^{\frac{1}{4\tilde{K}_{+}}}}
\end{equation}%
where we have replaced $\cos \left( \theta _{1,2-}\right) $ and $\cos (\sqrt{%
2}\Theta _{-})$ by their non-zero expectation values. Therefore, $\tilde{K}%
_{+}$ can be estimated numerically in the gapless phase by measuring the
power law decay of this correlation function. We conclude that the phase
penetrating between the HI and MI at finite $t_{\perp }$ is a SF phase which
remains stable as long as the above correlation function decays with a power
smaller than $\frac{1}{4}$.

In Ref. [\onlinecite{GiamarchiLadder}] it was found that turning on inter
chain tunneling between two chains that are in the MI phase can drive a
transition to a superfluid at a critical value of $t_\perp$. Here we showed
how this transition drops to $t_\perp=0$ as one approaches the MI-HI
critical point.

The actual value of the Luttinger parameter $\tilde{K}_{+}$ can be
calculated reliably only in the weak coupling regime. To estimate it for
strong coupling we use numerical DMRG simulations (Section \ref{sec:dmrg}).
If either the intra-chain $U$, $V$ or the inter-chain $V_{\perp }$ are
increased sufficiently, $\tilde{K}_{+}$ can be driven below the critical
value of $1$. Then the system would undergo a Kosterlitz-Thouless (KT)-type
transition to an insulating phase.

We finally note that in the limit of vanishing inter-chain tunneling $%
t_{\perp }=0$, but finite interaction $V_{\perp }\neq 0$, no intermediate SF
phase is formed near the MI to HI\ phase boundary. The inter-chain coupling
is in this case marginally irrelevant at the critical point, affecting only
a renormalization of the Luttinger parameter.

\emph{HI to DW transition -- } This transition is of the Ising universality
class\cite{Schulz,numerics_xxz}. It is controlled by the odd sector single
chain Hamiltonian $H_{-}$ [Eq. (\ref{Hm})], which describes a competition
between the terms $\cos (2\theta _{-})$ and $\cos (2\phi_{-})$. The
transition to the DW phase occurs when the latter term becomes more
relevant, which for small $g_{2}$, $g_{3}$ happens at $K_{-}=1$. For this
value of $K_-$ the theory (\ref{Hm}) can be refermionized and written as a
quadratic Hamiltonian of Dirac fermions with a mass term proportional to $%
g_{2}$ and a pairing term proportional to $g_{3}$. The model is diagonalized
when formulated in terms of two Majorana fields\cite{SheltonTsvelik}. At the
critical point $g_2=g_3$ (and $K_-=1$), one of these Majorana fields becomes
massless. A massless Majorana theory is equivalent to an Ising critical
point.

$H_{t_{\perp }}$ is clearly irrelevant at the critical point, since it
contains $\cos \left( \sqrt{2}\Theta _{-}\right) $ while the dual field $%
\Phi _{-}$ is pinned. The only effect this term could have is a slight
bending of the HI\ to DW\ phase boundary.
\begin{figure*}[t]
\includegraphics[scale=0.8]{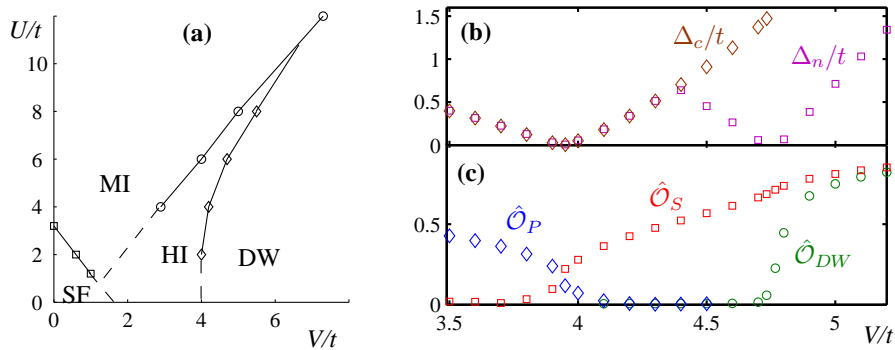}\centering
\caption{(Color online.) (a) Phase diagram of a single unperturbed chain in the space of $%
(U/t,V/t)$, obtained from the numerical calculations. (b) The
``charge" ($\diamond$) and ``neutral" ($\square$) gaps along the
same cut through the three insulating phases. The charge gap
vanishes at the MI-HI transition, whereas only the ``neutral" gap
vanishes at the transition to the DW phase. (c) The parity
($\diamond$), string ($\square$) and Density wave (o) order
parameters as a function of $V/t$ along a line of constant
$U/t=6$. The string and parity orders are defined as the square
roots of (\protect\ref{Cs}) and (\protect\ref{Cp}) respectively.}
\label{fig:chain}
\end{figure*}

On the other hand we show below that $H_{V_{\perp }}$ is a relevant
perturbation at the Ising critical point. First we can replace the operator $%
\cos (\sqrt{2}\Phi _{-})-\cos (\sqrt{2}\Phi _{+})$ appearing in $H_{V_{\perp
}}$ by its non-zero expectation value. (Note that these operators involve
only the even modes $\phi _{\alpha ,+}$ of each chain, and the even sector
Hamiltonian is not close to its critical point.) Now in order to determine
the scaling dimension of $H_{V_{\perp }}$, we need the scaling dimension of
the operators $\cos \left( \phi _{1/2,-}\right) $ at the critical point. As
mentioned above, there is a correspondence, via the mapping to a the
Majorana theory, between these operators and the two Ising models that
describe the odd sector Hamiltonain close to the critical point. The $\cos
\left( \phi _{1/2,-}\right) $ operators correspond to products of the spin
operators of the two Ising models\cite{LutherPeschel,ShankarBosonization}.
However since only one of these Ising models becomes critical at the
transition from HI to DW the operator $\cos \left( \phi _{1/2,-}\right) $
has the scaling dimension of Ising spins in $1+1$ dimensions at criticality,
which is $1/8$. $H_{V_{\perp }}$ is thus strongly relevant. This can also be
understood simply by the fact that at the transition to the DW\ phase, the
density wave susceptibility of each chain diverges, and therefore the
density-density coupling between the two chains is strongly relevant. $%
H_{V_{\perp }}$ then drives a transition to a density ordered phase with the
order parameters of the two chains locked to each other. The high $V_{\perp
} $ phase is thus not distinct from the $V_{\perp }=0$ DW phase.

The HI-DW phase boundary at finite $V_{\perp }$ can be obtained using the
same scaling argument as we used before for the phase boundary between the
HI or MI and the large $t_{\perp }$ phases. This gives that near the $%
V_{\perp }=0$ critical point, the critical line is of the form $V_{\perp
,c}\propto (V_{c}-V)^{7/4}$, where $V_{c}$ is the critical value for the
transition of the single chain from the HI to the DW\ phase (which is $%
V_{c}=\left( U+3\pi t\right) /2$ in the weak coupling limit).


In this section we studied weak coupling between two chains. The main results are summarized in the phase diagram plotted in Fig. \ref{fig:bos_phase}. The case of
an infinite array of chains is not expected to yield
qualitatively different results in the weak inter-chain coupling limit.
The main difference would be that the intermediate superfluid state would
in this case support true long range order instead of power-law correlations.

\section{DMRG results}

\label{sec:dmrg}

In this section we present numerical results to support the predictions of
the field theoretical analysis from the previous sections. To this end we
compute the ground state and lowest excitations of the extended Bose Hubbard
model (\ref{EBHM1}) and (\ref{Hperp}), using the Density Matrix
Renormalization Group (DMRG)\cite{white}. The section is organized as
follows. We first treat the case of a single unperturbed chain (Sec. \ref%
{sec:dmrgA}) at filling of ${\bar n}=1$ boson per site. Expanding on the
results of Ref. [\onlinecite{DallaTorre}], we focus on the interplay of the
non local string and parity correlations near the transition between the MI
and HI phases. In Sec. \ref{sec:dmrgB} we investigate how a perturbation
that breaks the lattice inversion symmetry affects the MI-HI phase
transition. We find that such a perturbation gaps out the critical point and
thus eliminates the phase transition, in agreement with the field
theoretical prediction. In Sec. \ref{sec:dmrgC} we move on to address a
two-leg ladder with only interaction coupling between the two chains.
Finally in Sec. \ref{sec:dmrgD} we analyze the two leg ladder with tunnel
coupling. As expected we find that this coupling leads to an intermediate
superfluid phase between the Mott and Haldane insulators.

Before proceeding let us give the essential technical details of the
numerical calculations. To expand the domain of the HI phase we add to (\ref%
{EBHM1}) the next-nearest neighbor interaction $(V/8) \sum_i \delta n_i
\delta n_{i+2}$. The DMRG calculations are performed with open boundary
conditions, while keeping $m=250$ states per block. As usual in bosonic
problems, we also need to truncate the Fock space of site occupations. For
calculations presented in this section we allow the four occupation states $%
n=0,1,2,3$. Including one more occupation state per site had a negligible
effect on the results in a sample of representative calculations. The
maximum number of sites in the calculations was 256, including chains of
length $L=256$ and two leg ladders of $L=128$.

\subsection{Unperturbed single chain}

\label{sec:dmrgA}
\begin{figure}[t]
\includegraphics[scale=0.7]{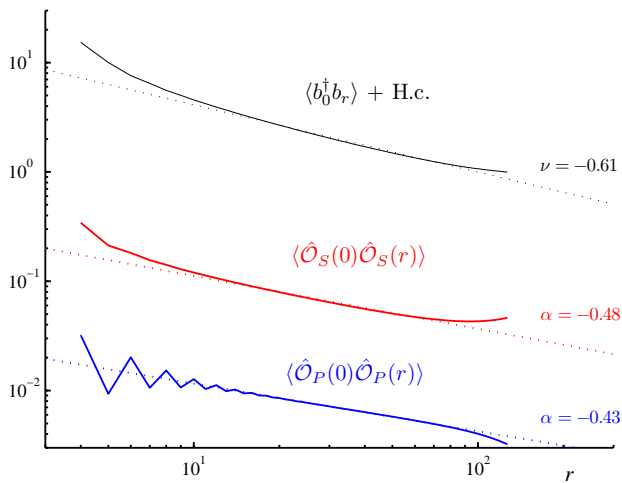}\centering
\caption{(Color online.) Decay of the parity and string
correlations, and of the single particle density matrix at the
HI-MI transition. The power-law fits are consistent with the field
theoretical predictions for the relations between the different
decay exponents (see text).} \label{fig:strings}
\end{figure}

To map the phase diagram of a single chain (Fig.
\ref{fig:chain}(a)) we compute the parity, string, and Density
wave (DW) correlations in the ground state. The long distance
behavior of these correlations as a function of $V$
for a particular value of the on-site interaction $U$ is plotted in Fig. \ref%
{fig:chain}(c). As expected, the MI is characterized by non
vanishing Parity ``order", the HI by string order, and the DW
phase by non decaying density correlations. The finite value of
the string and parity correlations seen in
the figure at the MI-HI critical point is due to the finite system size ($%
L=256$). The field theoretical analysis predicts power-law decay
of these correlation functions with a non universal power
${\alpha}$, which is directly related to the decay exponent $\nu$
of the single particle density matrix via the relation
${\alpha}=1/(4\nu)$. Along the line of critical points separating
the HI and MI phases the exponents are predicted to run in the
range $1/4<\nu<1$. These predictions are consistent with the
power-law fits of the relevant correlation functions at the
critical point, as presented in Fig. \ref{fig:strings} (See
explanation below on how the critical point is located in the
calculations).

Also shown in Fig. \ref{fig:chain}(a) is the superfluid (SF) phase
at low $U, V$, which is identified by measuring the decay exponent
of the single particle density matrix $\nu$. The SF phase is
stable when $\nu<1/4$.

In addition to ground state correlations we compute the gap to ``charged"
and ``neutral" excitations. The neutral excitation gap ${\Delta}_n$ is
obtained by targeting the lowest excitation in the sector with exactly ${%
\bar n}=1$ particles per site. The ``charge" gap is defined by ${\Delta}%
_c=E_0(+1)+E_0(-1)-2E_0(0)$, where $E_0(\pm 1)$ are the ground
state energies of the system with one more/less particle. There is
an interesting complication in extracting the bulk gap in the HI
phase. For open boundary conditions this phase supports low energy
edge excitations. We can identify these states by inspecting the
density profile of the wave functions. The appearance of the edge
states coincides with the transition to the HI phase and
facilitates the most precise determination of the transition point
in a finite system. In most cases, however, we are interested in
the bulk properties. To extract the bulk ``charge" and ``neutral"
gaps, we lift the edge excitations to high energy by applying a
sufficiently strong field at
the edges: $V_{edge}(\delta n_1 - \delta n_L)$. The gaps are plotted in Fig. %
\ref{fig:chain}(b) for a cut of the phase diagram at constant
$U/t=6$. It is seen that both the ``charge" and ``neutral" gaps
vanish at the transition from the MI to the HI phase. On the other
hand only the ``neutral" gap vanishes at the transition from the
HI to the DW phase.

\subsection{Breaking of lattice inversion symmetry}

\label{sec:dmrgB}
\begin{figure}[t]
\includegraphics[scale=0.7]{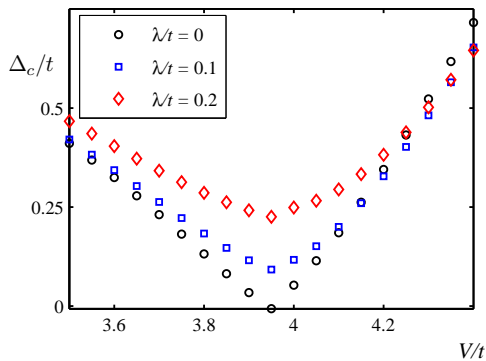}\centering
\caption{(Color online.) Charge gap as a function of $V/t$ at
constant $U/t=6$ for different values of the inversion-symmetry
breaking perturbation. $\protect\lambda$ is defined in
(\protect\ref{Hlambda}).} \label{fig:inv}
\end{figure}
Following the predictions of section \ref{sec:inv} we add the term
\begin{equation}
{\delta} H= \lambda\sum_i ({\delta} n_i b^\dagger_i b{^{\vphantom{\dagger}}}%
_{i+1}+ H.c.)  \label{Hlambda}
\end{equation}
to the Hamiltonian (\ref{EBHM1}). This is one of the simplest terms that
break the lattice inversion symmetry. Based on the predictions of \ref%
{sec:inv} we expect that the quantum critical point separating the HI and MI
phases would be eliminated in the presence of this term and an adiabatic
connection between the HI and MI would be facilitated. This is indeed what
is seen from the calculated charge gap in the presence of the perturbation.
In Fig. \ref{fig:inv} we plot the charge gap along a cut through the phase
diagram with $U/t=6$, as a function of $V/t$. We see that for a non
vanishing value of $\lambda$ the gap does not vanish and the phase
transition is eliminated.

\subsection{Two leg ladder with inter-chain repulsive interaction}

\label{sec:dmrgC}
\begin{figure}[t]
\includegraphics[scale=0.7]{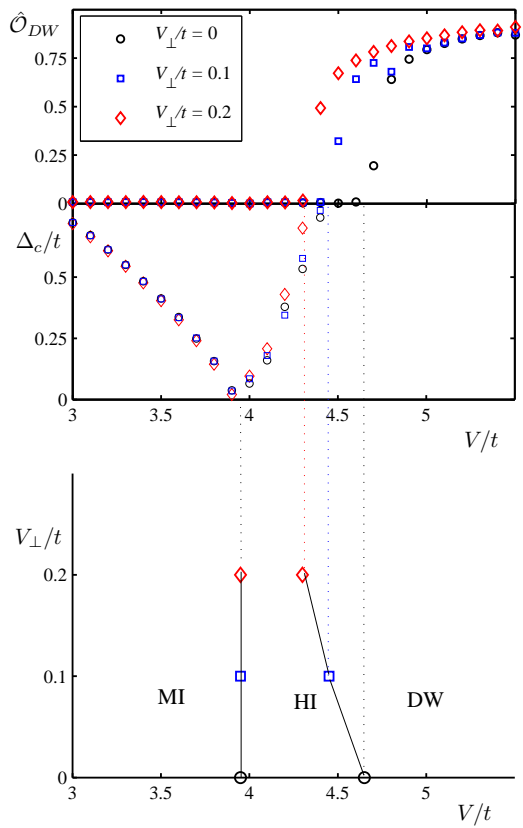}\centering
\caption{(Color online.) \emph{Effect of inter-chain interaction
coupling on the phase diagram.} (a) Calculated DW order parameter
as a function of $V/t$ at
constant $U/t=6$ for different values of the interchain interaction $V_\perp$%
. (b) Calculated charge gap on the same cut through the phase diagram and
the same values of $V_\perp$. (c) Evolution of the phase boundaries with $%
V_\perp$ inferred from the calculations.}
\label{fig:vperp}
\end{figure}

We move on to treat two leg ladders. The simplest coupling between two
chains is via the density-density interaction $V_\perp\sum_i n_{1,i} n_{2,i}$%
. The field theoretical analysis predicted that this interaction is marginal
at the transition between the MI and HI phases. In other words it leaves the
transition in tact, affecting only a renormalization of the Luttinger
parameter which controls the decay of the single particle density matrix at
the critical point.

At the transition from HI to DW the interaction coupling $V_\perp$
is 
expected to shift the critical point to lower values
of $V$. The change in both phase boundaries with increasing
$V_{\perp}$ are plotted in Fig. \ref{fig:vperp}. The transition
points are inferred from the vanishing of the ``charge" gap at the
MI-HI transition 
and the emergence of DW order
.

\subsection{Two leg ladder with Inter-chain tunnel coupling}

\label{sec:dmrgD}
\begin{figure}[t]
\includegraphics[scale=0.7]{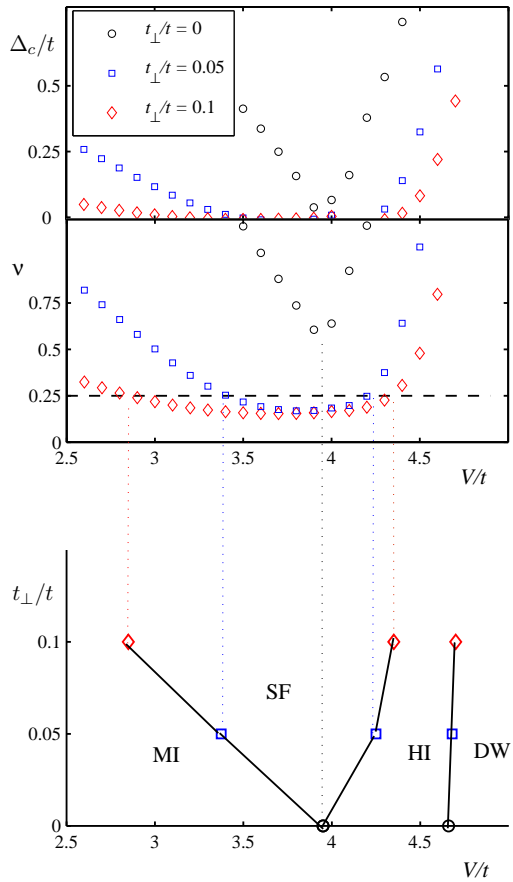}\centering
\caption{(Color online.) \emph{Effect of inter-chain tunnel
coupling on the phase diagram.} (a) Calculated charge gap as a
function of $V/t$ at constant $U/t=6$ for different values of the
inter-chain tunneling $t_\perp$. The gaps were obtained by
extrapolation to an infinite chain through finite size scaling
analysis. The fact that the estimate for the gap are sometimes
slightly negative is due to errors in the extrapolation. Note the
gapless phase that appears between the two insulating phases. (b)
Power-law fit for the spatial decay of the single particle density
matrix for the same cuts through the phase diagram. The fact that
${\nu}<1/4$ in the gapless phase confirms that it is indeed a
superfluid. The transition to the gapped phase seems to occur at
the universal exponent ${\nu}=1/4$ consistent with a
Kosterlitz-Thouless transition. (c) Evolution of the phase
boundaries with $t_\perp$ inferred from the calculations.}
\label{fig:tperp}
\end{figure}

Finally we include the inter-chain tunneling $- t_{\perp}
\left(b^\dagger_{A,i} b{^{\vphantom{\dagger}}}_{B,i} + \mathrm{H.c.}\right)$%
. According to the field theoretical analysis of the previous sections this
is a relevant coupling at the MI-HI critical point. An intermediate phase is
predicted to occur between these two insulators. In fact the intermediate
phase is expected to be a superfluid if the parameter $K_+ >1$. It would
seem that rather strong interactions are required to violate this criterion.
However we note that $K_+$ may be highly renormalized and not easily
estimated from the microscopic parameters. Therefore only the numerical
calculations presented in this section can confirm the nature of the
intermediate phase.

The results of the calculations are summarized in Fig. \ref{fig:tperp}.
The charge gap is shown as a function of $V/t$ in a cut through
the phase diagram at constant $U/t$. For increasing $t_\perp$ we
see a that a gapless phase opens up between the MI and HI phases.
Up to numerical accuracy and finite size effects the results are
consistent with the prediction that the gapless phase is
established for any non vanishing value of $t_\perp$.
The exponent ${\nu}$ with which we
fit the decay of the single particle correlation function is shown as a function of $V/t$. The fact that ${%
\nu}<1/4$ in the domain of vanishing gap is consistent with a
superfluid phase which is destabilized in a KT transition on
crossing to either the HI
or MI phases. We note that we find similar 
result also for different 
values of $U/t$ for which the HI phase can be realized.

\section{Summary and Discussion}

The purpose of this work was to investigate fundamental problems concerning
the nature of non local string order parameters, which were brought into
focus by the possible realization in systems of ultracold atoms\cite%
{DallaTorre}. In particular we addressed the question of how static
perturbations, which couple to \emph{local} physical operators can
nonetheless influence quantum phases and phase transitions that involve the
highly \emph{non-local} order parameter.

A key result of this analysis is the discovery of a surprising connection
between the string order and breaking of the lattice inversion symmetry
(Note that breaking of particle-hole symmetry is also needed -- but this is
anyway broken at the outset in the Bose system). In the context of this work
the two phases that are characterized by a non local order parameter are the
Haldane insulator, which sustains string order, and the Mott insulator,
which supports non zero expectation of the parity operator. We find that a
perturbation which breaks the lattice inversion symmetry gaps out the
critical point between the two phases. Correspondingly the sharp distinction
between them is eliminated, in the sense that non vanishing string order is
induced in the Mott phase and parity expectation value in the Haldane phase.
This is similar to the effect of a symmetry breaking perturbation on a
conventional phase transition involving spontaneous breaking of \emph{the
same} symmetry. But curiously, neither the Mott or Haldane phase involve
spontaneous breaking of the lattice inversion symmetry.

The seemingly mysterious connection between lattice inversion symmetry and
the string operators is elucidated by the effective long wavelength
description of the problem. The symmetry breaking field $\sin (2\phi_+)$ has
the symmetries of a local dipole field: it is odd under both lattice
inversion and particle-hole transformations. Incidentally, this operator can
also be decomposed as a product of the long wavelength expressions for the
string $\sin(\phi_+)$ and parity $\cos(\phi_+)$ operators. Thus in each
phase only one (non local) factor of the local ``dipole" field gains an
expectation value and so inversion symmetry remains intact. But this also
implies that in each of these phases the external perturbation becomes in
effect a direct coupling to the non local order parameter of the other
phase. For example in the MI phase where ${\langle \hat{\mathcal{O}}_P\rangle%
}={\langle \cos\phi_+\rangle}> 0$ we have that $\lambda\sin 2\phi_+ \approx
\lambda{\langle \hat{\mathcal{O}}_P\rangle}\hat{\mathcal{O}}_S$, which is
essentially an ordering field for the string order.

It is relatively easy to apply such a symmetry breaking field in an optical
lattice realization of the transition from Mott to Haldane insulator. The
principle has already been demonstrated successfully in experiments that
created a lattice of asymmetric double wells using a secondary laser with
half the wavelength of the main laser\cite{DoubleWells1,DoubleWells2}. A
straight forward extension would be to oscillate the secondary laser in
time. This can be used to measure the dynamical response of this system to
the local dipole field, an interesting theoretical problem, which we leave
for future work.

The second part of our analysis addressed the effect of coupling between
parallel one dimensional chains, which is inherent to realization with an
optical lattice. As a simple model for the coupled chains we analyzed the
case of two coupled chains, using a bosonization approach in section \ref%
{sec:weak2} and numerical simulations with DMRG in sections \ref{sec:dmrgC}
and \ref{sec:dmrgD}. The natural couplings to consider are inter chain
repulsion due to the dipolar moment of the atoms or molecules and the
inter-chain tunneling. The interaction coupling turns out to be essentially
trivial and does not change the structure of the phase diagram. The
inter-chain tunnel coupling, on the other hand, has a dramatic effect. Like
the inversion symmetry breaking perturbation discussed above, this
perturbation eliminates the distinction between the MI and HI phases.
However, instead of gapping out the critical point, it expands it into a
phase. More precisely, a weak inter chain tunneling gives rise to an
intermediate superfluid phase, whose domain grows quickly with the coupling
strength. The transition from the superfluid to either the Haldane or Mott
insulator is Kosterlitz-Thouless like and occurs at the universal value of
the decay exponent of the single particle density matrix.

The evolution of the phase diagram with increasing inter-chain coupling for
the case of many coupled chains and more general interactions deserves
further theoretical study. This may provide a controlled route for
investigating the crossover from one to two dimensions. Of particular
interest is the question of possible generalizations of the string orders to
the case of two (or at least quasi one) dimensional systems.

\emph{Acknowledgements.} We thank D. Arovas for useful
discussions, and D. Podolsky for his comments on this manuscript.
This research was supported in part by the Israeli Science
foundation and the U.S. Israel Binational Science Foundation (EA
and EGDT), by the Swiss NSF under MaNEP and Division II (TG) and
by D.O.E. grant \# DE-FG02-06ER46287 (EB). EA and TG are grateful
to the hospitality of the KITP Santa Barbara where parts of this
work were initiated and supported by NSF under Grant No.
PHY05-51164.
\appendix

\section{String correlations at weak inter-chain coupling}

\label{app:StringStability} In this appendix we derive the effect of weak
inter-chain coupling terms on the long range string correlations in the HI
phase. For simplicity we will study this within the effective spin-1 model
that results from restriction to three occupation states $n=0,1,2$ on each
site\cite{AltmanAuerbach2002,Huber}:
\begin{eqnarray}
H_{eff} &=&-{\frac{t{\bar{n}}}{2}}\sum_{{\langle ij\rangle}}\left(
S_{i}^{+}S_{j}^{-}+H.c.\right) +{\frac{U}{2}} \sum_{i}(S_{i}^{z})^{2}  \notag
\\
&&+V\sum_{{\langle ij\rangle }}S_{i}^{z}S_{j}^{z}-\mu\sum_i S^z_i  \notag \\
&&-{t{\bar{n}}\xi }\sum_{{\langle ij\rangle }}\left(
S_{i}^{-}(S_{i}^{z}+S_{j}^{z})S_{j}^{+}+H.c.\right)  \notag \\
&&-{\frac{t{\bar{n}}\xi ^{2}}{2}}\left(
S_{i}^{z}S_{i}^{+}S_{j}^{-}S_{j}^{z}+S_{i}^{+}S_{i}^{z}S_{j}^{z}S_{j}^{-}%
\right)  \label{Hs_full}
\end{eqnarray}%
where $\xi =\sqrt{2}-1$ for a system single average occupation. Here we
shall set $\xi =0$, essentially neglecting terms that break the particle
hole symmetry. Possible effects of these terms, which deserve further study,
will be briefly discussed at the end of this section.

Next we follow Kennedy and Tasaki and define a non local unitary
transformation of the spins on all lattice sites
\begin{equation}
U=\prod_{jk|j<k} e^{i\pi S^z_j S^x_k}
\end{equation}
The Hamiltonian gains an unusual, but nonetheless local form, in terms of
the transformed spin variables:
\begin{eqnarray}
\tilde{H} &=&-J\sum_{j} {\tilde{S}}^x_{j} {\tilde{S}}^x_{j+1} - {\tilde{S}}%
^y_{j} \exp(i \pi {\tilde{S}}^z_j + i \pi {\tilde{S}}^x_{j+1}) {\tilde{S}}%
^y_{j+1}  \notag \\
&&- V\sum_{j} {{\tilde{S}}}^{z}_j {{\tilde{S}}}^{z}_{j+1} + {\frac{U}{2}}
\sum_j ( {{\tilde{S}}}^{z}_j)^2 .  \label{Htilde}
\end{eqnarray}
$\tilde H$ has an explicit $Z_2\times Z_2$ symmetry, generated by $\pi$
rotations of the transformed spins around the main axes. Furthermore the non
local string correlations of the original spins map to standard two point
correlations of the transformed spins. In the Haldane phase the $Z_2\times
Z_2$ symmetry is broken, and the transformed spin operators gain a finite
expectation value.

We can therefore treat the Haldane phase of $\tilde H$ with a mean field
approximation\cite{KennedyTasaki}, in which the four degenerate broken
symmetry ground states are given by the product wavefunctions:
\begin{equation}
{\,|\,\Psi_{\zeta,{\eta}}\,\rangle\,}=\prod_i\left(\cos{\theta}{%
\,|\,0\,\rangle\,}_i+\zeta\sin{\theta}{\,|\,{\eta} 1\,\rangle\,}_i\right).
\end{equation}
The labels $\zeta,{\eta}=\pm 1$ correspond to the signs of the two $Z_2$
order parameters ${\langle \tilde S^x_i\rangle}$, and ${\langle \tilde
S^z_i\rangle}$. The elementary excitations in this phase are domain walls
separating any two of the degenerate ground states. Interestingly such
domain walls are generated by acting on a ground state with the original
spin operators:
\begin{eqnarray}
S^x_i&=&{\tilde S}^x_i \exp({i\pi\sum_{j>i}{\tilde S^x_j}})  \notag \\
S^y_i&=&\exp({i\pi\sum_{j<i}{\tilde S^z_j}}){\tilde S}^y_i \exp({i\pi\sum_{j>i}{%
\tilde S^x_j}})  \notag \\
S^z_i&=&\exp({i\pi\sum_{j<i}{\tilde S^z_j}}){\tilde S}^z_i
\end{eqnarray}
Thus $S^x_i$ creates a kink in ${\eta}$ (${\langle \tilde S^z_i\rangle}$), $%
S^z_i$ creates a kink in $\zeta$ (${\langle \tilde S^x_i\rangle}$), and $%
S^y_i$ creates kinks in both order parameters. As in any Ising system,
proliferation of a finite density of domain walls leads to destruction of
the order. This is the mechanism by which inter-chain coupling destroys the
long range string correlations.

Consider first the perturbative effect of the inter-chain interaction $%
H_{V_\perp}=V_\perp\sum_i S^z_{1i} S^z_{2i}$. The correction to
the ground state to first order in $V_\perp/\Delta$ consists of a
single domain wall in the ${\langle \tilde S^x_i\rangle}$ order
parameter in each chain. If we consider higher order terms, the
weight of a configuration with $n$ domain walls per chain scales
as $(V_\perp/\Delta)^n$ and the density of domain walls of
${\langle \tilde S^x_i\rangle}$ in the ground state is $\propto
V_\perp/\Delta$. Therefore we expect the $S^x$ string correlations
to decay exponentially over a correlation length
$\xi_x\sim\Delta/V_\perp$. On the other hand the $S^z$ string
order is left unmodified by this perturbation.

We can apply a similar argument to the inter-chain tunneling term $%
H_{t_\perp}=-t_\perp\sum_i (S^x_i S^x_j +S^y_i S^y_j)$.

\section{Alternative bosonization scheme}

\label{app:bos}

In section \ref{sec:weak} we derived the continuum field theory in a rather
indirect way. In the first step the EBHM (\ref{EBHM1}) was mapped to the
effective spin-1 model \ref{Hs_full}. We then followed the procedure
developed in Refs. [\onlinecite{TimonenLuther,Schulz}] for such spin systems
which involves: (i) splitting a spin-1 chain into two effective spin-$1/2$
chains; (ii) fermionizing the parallel spin-$1/2$ chains with a
Jordan-Wigner transformation; (iii) Bosonizing the fermions.

There is a standard procedure to ``bosonize" a bosonic Hamiltonian\cite%
{HaldaneBos,GiamarchiBook}, and it is tempting to ask why not to use this
more direct approach to derive the field theory which describes the
transitions between the MI, HI and DW phases. This turns out to be not so
trivial.

The standard scheme is based on expansion of the fluctuations of the
discrete particle density to slow modes describing small deviations from the
fundamental period set by the average density and its harmonics $k_n= 2\pi n
\rho_0$:
\begin{equation}
\rho(x) \to \left(\rho_0 - {\frac{1}{\pi}} \partial_x
\phi(x)\right)\sum_{n\in\mathbb{Z}} e^{{i(k_n x_i -2 n\phi(x_i))}}
\label{rhobos}
\end{equation}
Thus, slow variations around the fourier component $k=nk_0$ of the density
are given by the field $\cos\left(2n\phi(x)\right)$. A finite expectation
value for this field implies a static density wave, with the periodicity $%
\lambda_n= 2\pi/k_n=1/(n\rho_0)$ in the ground state. Clearly the period $2a$
density generated at unity filling by a strong nearest-neighbor interaction $%
V$ is not captured in this expansion (here $\rho_0=1/a$, with $a$ the
lattice spacing). It is therefore not surprising that the Haldane insulator,
which involves fluctuations at the same length scale and is in some sense a
precursor of the DW phase, cannot be described in this approach either.

We now propose a modified bosonization procedure that will enable us to
derive the low energy field theory (\ref{bosH1}). The key idea is to split a
bosonic chain into two auxiliary chains without changing the total number of
bosons. In other words, a single chain with one boson per site maps to a
ladder with one boson per rung, or one boson in every two sites in each of
the auxiliary chains. Next each of the half filled chains is bosonized
separately in the standard way
\begin{eqnarray}
b_{{\alpha }i} &=&e^{i\theta _{\a}(x_{i})}\left(
{\frac{1}{2a}}-{\frac{1}{\pi
}}\partial _{x}\phi _{\a}(x_{i})\right) ^{1/2}  \notag \\
&&\times \sum_{m\in \mathbb{Z}}e^{{i({\frac{m\pi }{a}}x_{i}-2m\phi
_{\a}(x_{i}))}}  \label{bos2} \\
{\delta }\rho _{\a}(x_{i}) &=&\left( {\frac{1}{2a}}-{\frac{1}{\pi
}}\partial
_{x}\phi _{\a}(x_{i})\right) \sum_{m\in \mathbb{Z}}e^{{i({\frac{m\pi }{a}}%
x_{i}-2m\phi _{\a}(x_{i}))}}  \notag
\end{eqnarray}%
where ${\alpha }=1,2$ is the auxiliary chain index. Note that the correct
wave-vector to describe the DW has emerged from the inverse density of each
of the split chains. We anticipate that the density wave of the physical
chain will appear as in-phase locking of density waves in the two auxiliary
half filled chains.

We now map the EBHM (\ref{EBHM1}) to a closely related model on the
auxiliary ladder by taking $b_i \to (b_{1i}+b_{2i})/\sqrt{2}$ and $n_i\to
n_{1i}+n_{2i}$. Choosing a slightly different extension of the BHM to the
auxiliary ladder system should not change the essential structure of the
phase diagram. The naive continuum limit of this model can now be taken by
using the identities (\ref{bos2}). For the on-site interaction term we
obtain:
\begin{widetext}
\be
\frac{U}2 \sum_i (n_{1,i} + n_{2,i} - 1)^2 \approx  \frac{Ua}2\int dx \Big[ \frac1{\pi^2} (\partial\phi_+)^2 -\frac{1}{a^2} \cos(2\phi_+)\cos(2\phi_+)
  + \frac1{a^2}\big(\cos(2\phi_+)+\cos(2\phi_-)\Big)\Big]\ee
where $\phi_\pm \equiv \phi_1 \pm \phi_2 $, and we have kept only the most relevant terms.
Similarly the nearest neighbor interaction term leads to
\be
{V\over 2}\sum_i (n_{1,i} + n_{2,i} - 1)(n_{1,i} + n_{2,i} - 1)
\approx Va\int dx \Big[ \frac1{\pi^2} (\partial\phi_+)^2 - \frac1{a^2}\cos(2\phi_+)\cos(2\phi_-) -\frac1{a^2}\Big(\cos(2\phi_+)+\cos(2\phi_-)\Big)\Big]
\ee
Finally the hopping term translates to
\bea
 {t\over 2}\sum_i\sum_{\a,\b=1}^2 b\yd_{\a i}b\nd_{\b i+1}+H.c. &\approx& \frac{t}{2} \int dx \left[\left(a(\partial\theta_+)^2 + a(\partial\theta_-)^2 \right)\left(1+\cos(2\theta_-)\right) - \frac{4}a\cos(2\phi_+)\cos(2\phi_-) (1+2\cos(2\theta_-))\right. \nn\\
&&+\left. \left( - \frac{a}{4\pi^2}(\partial\phi_-)^2+ \frac4{a}\cos(2\phi_+) + \frac4{a}\cos(2\phi_-)- \frac{2}{a}\right)\cos(2\theta_-)\right].
\eea
\end{widetext}
Here ${\theta}_\pm\equiv ({\theta}_1\pm{\theta}_2)/2$. Note that we have
essentially the same degrees of freedom here as in section \ref{sec:weak1}.
In the MI and HI phases the operator $\cos(2{\theta}_-)$ is relevant and we
can safely replace it with its expectation value $C_-$ which is of order 1.
In this case the above expressions simplify and give precisely the field
theory (\ref{bosH1}) with the parameters:
\begin{eqnarray}
K_+ &=& \pi\sqrt{\frac{\frac{t}2(1+C_-)}{\frac{U}2+V}},~~K_- = 2\pi\sqrt{%
\frac{1+C_-}{C_-}}  \notag \\
g_1 &=& g_2 =\pi^2 a\left(\frac{U}2 -V - 2tC_-\right),~~g_3 = -\pi a t
\notag \\
g_4&=&-\pi^2 a\left[\frac{U}{2} +V + 2t\left(1+2C_-\right)\right]
\end{eqnarray}
Note that the bare value of the Luttinger parameter $K_-$ of the
antisymmetric fields is large ($K_- >2\pi$), consistent with taking $C_-$ of
order 1.

Before closing this appendix let us make a few remarks concerning the new
scheme. First note that the bare value of the Luttinger parameter $K_{+}$
implies a reasonable estimate for the phase boundary between the superfluid
and insulating phases. The system should be superfluid when the renormalized
Luttinger parameter ${\tilde{K}}_{+}$ exceeds 2. Estimating this criterion
with the bare value given above we get the approximate criterion $%
(U+2V)/t<\pi ^{2}/4$. By contrast the spin-1 mapping predicted that there is
no superfluid phase in the region $U>0$ and $V>0$. Indeed we do not expect
the spin-1 mapping, which involves truncation to 3 occupation states, to
hold when the on-site repulsion is not sufficiently strong. Thus the new
scheme improves on the the spin one mapping in that it gives reasonable
predictions for the phase diagram already at the level of the naive
continuum limit.

Another notable difference between the field theory of section \ref%
{sec:weak1} and the new scheme is a $\pi $ shift in the definition of the
field $\phi _{+}$. Hence in the bosonic scheme $\phi _{+}\approx \pi $
corresponds to the MI phase whereas $\phi _{+}\approx 0$ to the HI. This is
consistent with the physical interpretation of the variable $\phi _{+}$ in
the new scheme. Consider first the DW phase, which corresponds to a phase
locked density wave on the two auxiliary chains. For this we need $\phi _{1}$%
=$\phi _{2}\approx 0$. Since $\phi _{+}$ is not critical at the transition
from DW to HI we expect $\phi _{+}\approx 0$ also in the HI phase.

\section{the String and Parity order parameters}

\label{app:BosString}

The HI and MI phases do not support long range order in any local order
parameter. Instead, they are characterized by the non-local string and
parity order parameters (Eqs. (\ref{Os},\ref{Op}), respectively). In order
to obtain the bosonized expressions for these order parameters in the
effective field theory, we need to find their continuum limit. Here we
propose a continuum form of these operators using general considerations
based on the asymptotic \textquotedblleft particle-hole" symmetry of the
model at low energy. More microscopic derivations can be found in Refs. [%
\onlinecite{GiamarchiString,Nakamura}]

In the following argument, we will assume that it is legitimate to truncate
the Hilbert space to the three lowest occupation numbers ($\delta
n_{j}=0,\pm 1$). $\delta n$ can then be represented by a pseudospin-1 degree
of freedom, defined by $S_{j}^{z}=\delta n_{j}$. The model (\ref{EBHM1}) is
then replaced by the effective spin-1 model of Eq. (\ref{H_s}), which leads
to the same low energy effective theory (\ref{Hp},\ref{Hm}). We have
neglected terms that break the symmetry between $S_{j}^{z}=-1 $ and $%
S_{j}^{z}=1$ (``particle-hole" symmetry). These terms appear in Eq. (\ref%
{Hs_full}). They are irrelevant at low energies, so neglecting
them should not change the long-distance behavior.

%
%

Next, we need to find the continuum limit of the operators $\hat{\mathcal{O}}%
_{P}\left( j\right) $ and $\hat{\mathcal{O}}_{S}\left( j\right) $. Taking
the naive continuum limit of $\hat{\mathcal{O}}_{P}\left( j\right) $, we get
$\hat{\mathcal{O}}_{P}\sim e^{i\phi _{+}}$ (since $\sum_{j<i}S_{j}^{z}%
\rightarrow \int^{x}dx^{\prime }S^{z}\left( x^{\prime }\right) =\frac{1}{\pi
}\phi _{+}\left( x\right) $, assuming that $\phi _{+}\left( -\infty \right)
=0$)\cite{Nakamura,ScalapinoString,NussinovString}. To get a hermitian
operator, $e^{i\phi _{+}}$ has to be symmetrized. To find the correct
symmetrization, we note that\ $\hat{\mathcal{O}}_{P}\left( j\right) $ should
be symmetric under a \textquotedblleft particle-hole" transformation (which
correspond under bosonization to $\phi _{+}\rightarrow -\phi _{+}$).
Therefore, $\hat{\mathcal{O}}_{P}$ should have the form
\begin{equation}
\hat{\mathcal{O}}_{P}\sim A_{P}\cos \left( \phi _{+}\right) +...
\label{O_p_app}
\end{equation}
where $A_{P}$ is a non-universal constant, and we have truncated additional
sub-leading operators of this sum. $\hat{\mathcal{O}}_{S}\left( j\right) $
also contains the same $e^{i\phi _{+}}$ factor, but it should be
anti-symmetric under a particle-hole transformation, which suggests the
general form
\begin{equation}
\hat{\mathcal{O}}_{S}=A_{S}\sin \left( \phi _{+}\right)
+B_{S}\partial _{x}\phi _{+}\cos (\phi _{+})+...  \label{O_s_app}
\end{equation}%

The second contribution in Eq. (\ref{O_s_app}) is expected from
naively bosonizing Eq. (\ref{Os}) and taking Eq. (\ref{O_p_app})
into account. However, a more careful treatment of the operator
product expansion for this expression
[\onlinecite{GiamarchiString,Nakamura}] shows that the first term
is also present.
The $\sin(\phi_+)$ term is the most relevant one. From Eqs. (%
\ref{O_p_app},\ref{O_s_app}) we see that in the MI phase, where
$\phi _{+}$ is pinned around $0$, we expect that $\langle
\hat{\mathcal{O}}_{P}\rangle \neq 0$, $\langle
\hat{\mathcal{O}}_{S}\rangle =0$, while in the HI $\phi
_{+}$ is pinned around $\pm \pi /2$, therefore $\langle \hat{\mathcal{O}}%
_{P}\rangle =0$, $\langle \hat{\mathcal{O}}_{S}\rangle \neq 0$. In order to
test the validity of these results for the EBHM [Eq. (\ref{EBHM1})], where
particle-hole breaking terms exist, we evaluate the string and parity
correlation functions in (\ref{Cs},\ref{Cp}) numerically across the MI$%
\rightarrow $HI transition. The results, summarized in Figs. \ref{fig:chain},%
\ref{fig:strings}, are consistent with Eqs. (\ref{O_p_app},\ref{O_s_app}).



\begin{thebibliography}{99}
\bibitem{FootnoteFinite} In a finite system with open boundary conditions,
the system can break inversion symmetry due to the existence of edge states
in the HI phase.
\end{thebibliography}
\end{document}